\documentclass[aps,prd,nofootinbib,longbibliography,a4paper,11pt,eqsecnum]{revtex4-1}
\usepackage{amsmath} 
\usepackage{graphicx} 
\usepackage{amsthm}
\usepackage{amssymb} 
\usepackage{dsfont}
\usepackage{yfonts}
\usepackage{hyperref}
\usepackage{array,xcolor,graphicx}
\usepackage{booktabs,multirow}
\usepackage[utf8]{inputenc}
\usepackage{mathtools}

\usepackage{etoolbox}
\patchcmd{\section}
{\centering}
{\raggedright}
{}
{}
\patchcmd{\subsection}
{\centering}
{\raggedright}
{}
{}
\usepackage[all]{xy}
\usepackage{tikz}
\usetikzlibrary{arrows.meta}

\hypersetup{colorlinks=true,linkcolor=blue,citecolor=red,urlcolor=blue}

\newcommand{\be}{\begin{equation}}
\newcommand{\ee}{\end{equation}}
\newcommand{\bea}{\begin{eqnarray}}
\newcommand{\eea}{\end{eqnarray}}

\begin{document}
	
	\title{Lifshitz hydrodynamics at generic {\it z\/} from a moving black brane}
	
	\author{Aruna Rajagopal}
	\author{Larus Thorlacius}
\affiliation{University of Iceland, Science Institute, Dunhaga 3, IS-107, Reykjavik, Iceland}

	\begin{abstract}
	\vspace{1cm}
	A Lifshitz black brane at generic dynamical critical exponent $z>1$, with non-zero linear momentum along the boundary, provides a holographic dual description of a non-equilibrium steady state in a quantum critical fluid, with Lifshitz scale invariance but without boost symmetry. We consider moving Lifshitz branes in Einstein-Maxwell-Dilaton gravity and obtain the non-relativistic stress tensor complex of the dual field theory via a suitable holographic renormalisation procedure. The resulting black brane hydrodynamics and thermodynamics are a concrete holographic realization of a Lifshitz perfect fluid with a generic dynamical critical exponent. 
	\end{abstract}
	\maketitle
	\begingroup
	\hypersetup{linkcolor=black}
	\tableofcontents
	\endgroup
	\hrulefill

\section{Introduction}

Fluid dynamics \cite{LLfluid} efficiently captures the long-wavelength, low frequency behavior of a large class of interacting physical systems at finite temperature. Based on general symmetry principles and the conservation of local currents, it is a universal theory, as even systems with varying microscopic dynamics, can have the same macroscopic, hydrodynamic description. 
An appropriate hydrodynamic model, incorporating the dissipative effects of the thermal medium that are essential for a fluid to equilibriate after being perturbed away from equilibrium, consists of a gradient expansion of the conserved currents. At a given order, the conservation equations determine the expansion up to a finite number of undetermined coefficients. These coefficients may then be obtained either from measurements or from microscopic computations. For strongly coupled quantum systems, explicit computations are difficult and a gauge/gravity formulation \cite{Policastro_2001,Policastro_2002,Policastro_2002b} can offer advantages.
Einstein’s equations with a negative cosmological constant, supplemented with appropriate regularity restrictions and boundary conditions, reduce to the nonlinear equations of fluid dynamics in an appropriate parameter range, and a systematic framework to construct this universal nonlinear fluid dynamics, order by order in a boundary derivative expansion has been developed \cite{Bhattacharyya:2008jc, Baier:2007ix}.

In standard anti-de Sitter (AdS) holography \cite{Maldacena_1997} the dual field theory is relativistic and much of the dual gravitational formulation of hydrodynamics has focused on relativistic systems. However, many strongly correlated systems in nature are inherently non-relativistic and this has motivated the application of holography to a wider setting involving gravitational bulk theories with asymptotic geometries that are not AdS \cite{Son_2008, Balasubramanian_2008,Kachru_2008}. Generic non-relativistic quantum critical fluids have an emergent scaling symmetry of the form
\be\label{lifshitzscaling}
t\rightarrow \Lambda^z\,t,\qquad \vec{x}\rightarrow \Lambda \vec{x},
\ee
commonly referred to as Lifshitz scaling and are characterised by a dynamical critical exponent $z\geq 1$. For $z>1$, the scaling is asymmetric between the time and spatial directions and the system has scale invariance without conformal invariance. By now, there is a large body of work on holographic systems that realize Lifshitz scaling of the form \eqref{lifshitzscaling} at the asymptotic boundary of the gravitational bulk space-time, including a number of papers with a specific focus on hydrodynamic aspects, see {\it e.g.} \cite{Taylor:2008tg, Ross:2009ar, Ross_2011, Mann:2011hg, Bertoldi:2009dt, Hoyos:2013eza, Hoyos:2013qna,Kiritsis:2015doa, Hartong_2016}. 

Quantum critical fluids with Lifshitz symmetry, or Lifshitz fluids for short, have interesting properties that set them aside from more conventional fluid systems. In particular, a Lifshitz fluid with generic dynamical critical exponent $z\neq 1,2$ cannot have boost symmetry \cite{de_Boer_2018, Grinstein_2018}. A general formalism for perfect fluids without boost symmetry has been developed in \cite{de_Boer_2018, de_Boer_2020, Novak:2019wqg} and a primary motivation for the present paper is to provide a concrete holographic setup for the study of such fluids.

In previous work \cite{Fern_ndez_2019}, we have studied out-of-equilibrium energy transport in a quantum critical fluid with Lifshitz scaling symmetry following a local quench between two semi-infinite fluid reservoirs. The late time energy flow was found to be universal and was accommodated via a steady state occupying an expanding central region between outgoing shock and rarefaction waves, in accordance with earlier work on critical fluids with Lorentz boost symmetry \cite{Bhaseen:2013ypa, Lucas_2016b, Spillane_2016,Amado:2015uza, Pourhasan_2016}. In the relativistic case, a holographic version of the quench problem was considered in \cite{Bhaseen:2013ypa,Amado:2015uza} and it would be interesting to extend this to Lifshitz fluids without boost symmetry. As a first step in that direction, we consider a Lifshitz black brane at generic $z$, moving with uniform velocity $v$ in one of the transverse directions, as a holographic dual description of a non-equilibrium steady state of the corresponding Lifshitz fluid. 

The holographic model we choose to work with is a relatively simple variant of Einstein-Maxwell-Dilaton gravity with exact analytic solutions describing static Lifshitz black branes. The model can easily be generalised to include charged matter in order to study holographic duality at finite temperature and chemical potential. The exact static black brane solution provides a convenient benchmark to test our formalism against, even if the moving brane solution is beyond exact analytic control. A drawback of this model, that has detracted from its usefulness for holography in the past, is the fact that the asymptotic Lifshitz symmetry is obscured by a running dilaton and non-standard asymptotic behaviour of the background vector field. It turns out, this can easily be remedied by taking advantage of a built-in symmetry of the model and a more or less standard holographic renormalisation is achieved by suitably adapting the formalism developed for a holographic model involving a massive vector field in \cite{Ross:2009ar}. 
We find the following results:
\begin{itemize}
    \item 
    A physically distinct class of moving Lifshitz black brane solutions in four spacetime dimensions, dual to a uniformly moving 2+1 dimensional Lifshitz fluid. The fluid velocity plays the role of chemical potential, dual to the momentum density, in line with the perfect fluid formalism developed in \cite{de_Boer_2018,de_Boer_2020}.
    \item 
    A renormalised boundary stress tensor complex whose components express the energy density $\mathcal{E}$, energy flux $\mathcal{E}^i$, momentum density $\mathcal{P}_i$, and spatial stress tensor $\Pi_{ij}$ of the dual Lifshitz fluid.
    \item  
    The Ward identity associated to Lifshitz scaling leads to the equation of state $z\mathcal{E} - \rho v^2 = d_s P$, where $d_s=2$ is the number of transverse dimensions. This is precisely the equation of state hypothesised for a perfect Lifshitz fluid with an arbitrary dynamic exponent $z$ in \citep{Hartong_2016,de_Boer_2018}.
\end{itemize}

\section{The gravitational theory}
\label{bulktheory}

We restrict our attention to a holographic theory with Lifshitz scaling defined in 4 bulk space-time dimensions but our results can easily be generalised to an arbitrary number of dimensions. The model we work with is a simple variant of Einstein-Maxwell-Dilaton (EMD) theory, which consists of Einstein gravity along with a massless $U(1)$ gauge field, $A_{\mu}$ and a dilaton field, $\phi$. Using a normalisation where $16\pi G_4 = 1$, the action is given by,
\begin{equation}
S_0 = \int d^4x \sqrt{-g}\left[R-2\Lambda-\frac{1}{2}(\partial\phi)^2-\frac{1}{4}e^{\lambda\phi}F^2\right]+ 2\int d^3x \sqrt{-\gamma}K,
\label{model}
\end{equation}
where the last term is the usual Gibbons-Hawking-York boundary term, which is needed in order to have a well-defined variational problem for the metric. 
The equations of motion for \eqref{model} are,

\begin{align}
R_{\mu\nu}-\frac{1}{2}R g_{\mu\nu} + \Lambda g_{\mu\nu} &= T_{\mu\nu}^{\phi}+T_{\mu\nu}^{(1)}\\
\nabla^2\phi - \frac{\lambda}{4}e^{\lambda \phi}F_{\mu\nu}F^{\mu\nu} &=0\\
\nabla_{\mu}e^{\lambda \phi}F^{\mu\nu} &=0,
\end{align}
with,
\begin{align}
T_{\mu\nu}^{\phi} &= \frac{1}{2}\partial_{\mu}\phi\partial_{\nu}\phi - \frac{1}{4}g_{\mu\nu}(\partial \phi)^2,\\
T_{\mu\nu}^{F} &= \frac{1}{2}e^{\lambda \phi}\left(F_{\mu\sigma}F^{\sigma}_{\nu}-\frac{1}{4}g_{\mu\nu}F_{\sigma\rho}F^{\sigma\rho}\right).
\end{align}

Models of this kind were introduced in the context of non-relativistic holography in \cite{Taylor:2008tg}. This is by no means the only possible model for Lifshitz holography but it has an important advantage in that there exists a full analytic solution to the field equations that describes a static black brane in asymptotically Lifshitz space-time. The model can easily be generalized to include charged black brane solutions that are Lifshitz analogs of  AdS-Reissner-Nordstr\"om black branes \cite{Tarr_o_2011}, which are key to a holographic dual description of non-relativistic quantum critical matter at finite temperature and chemical potential, see {\it e.g.} \cite{Keranen:2012mx}. The bulk theory can also include matter fields of various types, but this is not needed for the main purpose of this paper, which is to carry out holographic renormalisation for moving Lifshitz branes and establish thermodynamic relations that hold for the resulting Lifshitz hydrodynamics.

The model \eqref{model} admits the so-called Lifshitz space-time as a solution,
\begin{equation}\label{lifshitzmetric}
	ds^2 = \ell^2\left(-r^{2z}dt^2 + \frac{dr^2}{r^2} + r^2d\vec{x}^2 \right),
\end{equation}
where $\ell$ is a characteristic length scale of the geometry which we set to unity for notational simplicity. The metric exhibits the required Lifshitz scaling,
\begin{equation}
	t \rightarrow \Lambda^z t \ ,\qquad x \rightarrow \Lambda x\ ,\qquad r \rightarrow \Lambda^{-1}r,
	\label{lifscaling}
\end{equation}
with $z\geq 1$. Space-time geometries that are  asymptotic to this metric provide a holographic dual description of a scale-invariant non-relativistic field theory formulated on a $\mathbb{R}^t\times\mathbb{R}^2$ boundary.  

The metric \eqref{lifshitzmetric} is a solution to the equations of motion provided the parameters of the model satisfy $\lambda = -\frac{2}{\sqrt{z-1}}$ and  $\Lambda = -\frac{1}{2}(z+1)(z+2)$, and it is accompanied by a vector field and dilaton background of the form,
\begin{equation}\label{lifshitzbackground}
    A_{t} = \sqrt{\frac{2(z-1)}{z+2}}\left(\frac{r}{r_0}\right)^2r^{z}, \qquad e^{\phi} = \left(\frac{r}{r_0}\right)^{2\sqrt{z-1}},
\end{equation}
where $r_0$ is an arbitrary constant. 
While this model has the advantage of analytic control, it has the disadvantage of a logarithmically running dilaton and diverging vector field at the boundary. However, this is not a very serious disadvantage because the vector field only serves to provide the background to support a Lifshitz geometry at the boundary and does not couple to any non-gravitational fields.

Charged matter fields can be added to the model, but they should then be charged under some additional gauge field and not couple directly to the $A_\mu$ field considered here. The need for a further gauge field to accommodate charged matter can also be seen from the fact that the background field profile \eqref{lifshitzbackground} is not of the standard form that corresponds to having a chemical potential for an unbroken $U(1)$ symmetry in the dual field theory. 
In fact, $A_\mu$ should not be viewed as a gauge field but simply as a massless vector field that only interacts gravitationally (with a coupling that depends on the dilaton). With this in mind, we do not have to respect the $U(1)$ gauge symmetry of the bulk action in \eqref{model} when we construct boundary counterterms for holographic renormalisation of the model. 

The counterterms instead respect another symmetry. The action \eqref{model} is invariant under a constant shift of the dilaton field while simultaneously absorbing a constant normalisation factor into the gauge field, 
\begin{equation}\label{shiftsymmetry}
\phi \rightarrow \phi -\frac{2}{\lambda}\log\alpha,\qquad A_{\mu} \rightarrow \alpha A_{\mu},
\end{equation}
and this will be a symmetry of our boundary action as well. The shift symmetry is also helpful in analysing the asymptotic behaviour of bulk fields.

\subsection{Black branes with linear momentum}

We would like to construct the gravitational dual of a perfect Lifshitz fluid moving at non-vanishing velocity. We are particularly interested in fluids at generic $z$, which do not have boost symmetry. In this case, it is not enough to perform a boost of the black brane along one of the transverse directions and study the associated thermodynamics as this does not give us a genuine dual of a moving Lifshitz fluid, but instead corresponds to studying a fluid at rest from a moving coordinate frame. The way around this is to construct from scratch a bulk solution that describes a moving Lifshitz black brane at generic $z \neq 1$, with metric, dilaton and vector fields that encode the fluid momentum. 
For this we adopt an ansatz employed in \cite{Hartong_2016} for a $z=2$ moving brane and adapt it to the more general case. With the linear momentum taken to be in one of the transverse directions, say the $y$ direction, the metric becomes
\begin{equation}
ds^2 = -F_1(r)r^{2z}dt^2 + \frac{dr^2}{r^2 F_2(r)} + r^2 F_3(r)dx^2 
+ F_4(r)\big(rdy + N(r)r^z dt\big)^2,
\label{metric}
\end{equation}
and the vector field is 
\be
A = r^z G_1(r)dt + G_2(r)\big(rdy + N(r)r^z dt\big).
\label{gaugefield}
\ee
Note that this ansatz includes as a special case an exact solution for a static black brane by simply setting $G_2 = N = 0$, $F_3 = F_4=1$, and letting $F_1=F_2=1-(r_h/r)^{z+2}$. The dilaton field of the static black brane solution is unchanged from \eqref{lifshitzbackground} provided we use the shift symmetry \eqref{shiftsymmetry} to set $r_0$ equal to $r_h$, the radial location of the event horizon. 

There is no known analytic solution for a brane with non-vanishing linear momentum. The field equations can be solved numerically, by imposing appropriate boundary conditions at the black brane horizon, but we will not pursue a numerical solution here. The results we are after can instead be obtained by employing a combination of asymptotic analysis and radially conserved charges along the lines of \cite{Hartong_2016}. In the following subsections we first derive a set of Noether charges that are constant along the radial coordinate. We then introduce boundary counterterms and implement them in the holographic renormalisation of the boundary stress tensor. Finally, we solve the linearised field equations around the Lifshitz fixed point in \eqref{lifshitzmetric} and \eqref{lifshitzbackground}. The different elements are then brought together in Section~\ref{sect:thermo} where we analyse the thermodynamics of the moving black brane.

\subsection{Noether charges}
\label{noetherdef}

The fields that enter in the moving brane ansatz \eqref{metric} - \eqref{gaugefield} are functions of  the radial coordinate and in this case the action \eqref{model} can be reduced to an integral over a one-dimensional Lagrangian by factoring off the integrals over the time and transverse coordinates. The Gibbons-Hawking-York term cancels against boundary terms resulting from integration by parts of terms in the radial bulk action, leading to the result,
\begin{equation}\label{L1D}
\begin{split}
L_{1D} =&r^{z+1}\sqrt{F_1F_2F_3F_4}\Big[-2(z^2+2z+3)+\frac{1}{2}\frac{F_4N^2}{F_1}\Big(z-1+\frac{rN'}{N}\Big)^2-z\frac{rF_1'}{F_1}\\&-(z+2)\frac{rF_2'}{F_2}-\frac{rF_3'}{F_3}-\frac{rF_4'}{F_4}+\frac{r^2}{2}\Big(\frac{F_1'}{F_1}\frac{F_3'}{F_3}+\frac{F_1'}{F_1}\frac{F_4'}{F_4}+\frac{F_3'}{F_3}\frac{F_4'}{F_4}\Big)-\frac{2\Lambda}{F_2}-\frac{1}{2}(r\phi')^2\\&+\frac{1}{2}e^{\lambda\phi}\Big(\frac{1}{F_1}(zG_1+rG_1'+G_2((z-1)N+rN'))^2-\frac{1}{F_4}(G_2+rG_2')^2\Big)\Big],
\end{split}
\end{equation}
where $' \equiv \frac{d}{dr}$. The terms in the 1D Lagrangian are arranged so as to make apparent the following two scaling symmetries,
\begin{align}
&F_1 \rightarrow \alpha^2 F_1, \quad F_3 \rightarrow \alpha^{-1}F_3, \quad F_4 \rightarrow \alpha^{-1}F_4,\\
&N \rightarrow \alpha^{\frac{3}{2}}N, \quad G_1 \rightarrow \alpha G_1, \quad G_2 \rightarrow \alpha^{-\frac{1}{2}}G_2,
\end{align}
and
\begin{align}
&F_3 \rightarrow \beta^{2}F_3, \quad F_4 \rightarrow \beta^{-2}F_4,\\
&N \rightarrow \beta N,\quad G_2 \rightarrow \beta^{-1}G_2.
\end{align}
They represent diffeomorphisms which preserve the volume element $dtdxdy$, and can thus be thought of as a Noether symmetry inherited by $L_{1D}$. They are also symmetries of the boosted metric ansatz \eqref{metric}, as long as the coordinates transform as $t \rightarrow \alpha^{-1} t,\; x \rightarrow \alpha^{\frac{1}{2}} x,\; y \rightarrow \alpha^{\frac{1}{2}}y$ and $x \rightarrow \beta^{-1} x ,\; y \rightarrow \beta y$ respectively.

The two Noether charges associated with these symmetries are found to be,
\begin{equation}\label{Q1}
\begin{split}
    Q_{\alpha} &= r^{z+2}\sqrt{F_1F_2F_3F_4}\Big[2(z-1)+\frac{r}{2}\Big(\frac{2F_1'}{F_1}-\frac{F_3'}{F_3}-\frac{F_4'}{F_4}\Big)-\frac{3}{2}\frac{F_4N^2}{F_1}\Big(z-1+\frac{rN'}{N}\Big)\\&-\frac{1}{2}e^{\lambda\phi}\frac{G_2}{F_4}(G_2+rG_2')-e^{\lambda\phi}\Big(\frac{G_1}{F_1}+\frac{3}{2}\frac{G_2N}{F_1}\Big)\Big(zG_1+rG_1'+G_2N\Big(z-1+\frac{rN'}{N}\Big)\Big)\Big],
    \end{split}
   \end{equation}

\begin{equation}\label{Q2}
\begin{split}
    Q_{\beta} &= r^{z+2}\sqrt{F_1F_2F_3F_4}\Big[r\Big(\frac{F_3'}{F_3}-\frac{F_4'}{F_4}\Big)-\frac{F_4N^2}{F_1}\Big(z-1+\frac{rN'}{N}\Big)\\&- e^{\lambda\phi}\frac{G_2}{F_1}\Big(\frac{F_1}{F_4}(G_2+rG_2')+(zG_1+rG_1')N+G_2N^2\Big(z-1+\frac{rN'}{N}\Big)\Big]
    \end{split}
\end{equation}
These charges are combinations of bulk fields that do not depend on the radial bulk coordinate. This is particularly useful in relating horizon data to boundary data, and plays an important role in determining the thermodynamic equation of state when the exact bulk interpolating solution is not known, as we will see later on. 

There is another conserved charge in our system associated with the shift symmetry of the dilaton \eqref{shiftsymmetry}. While it does not play a direct role in the thermodynamics, it will be useful later on when we consider solutions of the linearized field equations. It is given by,
\begin{equation}
    Q_{\phi} = r^{z+2}\sqrt{F_1F_2F_3F_4}\Big[e^{\lambda\phi}\frac{G_2}{F_4}(G_2+rG_2') -e^{\lambda\phi}\frac{G_1}{F_1}\Big(zG_1+rG_1'+G_2N\Big(z-1+\frac{rN'}{N}\Big)-\frac{2r\phi'}{\lambda}\Big].
    \label{q3}
\end{equation}

In addition to the above symmetries, there is also a local gauge symmetry involving $r$ diffeomorphisms, which we fix by setting $F_3(r) = 1$ in \eqref{metric}.

\subsection{Holographic renormalisation}
In this section, we work out the renormalised stress-energy tensor at the boundary for the EMD model. The conservation of the stress tensor gives us the standard energy and momentum conservation equations, while its scaling behaviour under a Lifshitz transformation will give us the equation of state of a Lifshitz fluid. For holographic renormalisation of gravity models in asymptotically Lifshitz spacetime see {\it e.g.} \cite{Ross:2009ar, Ross_2011, Zingg:2011cw, Baggio:2011cp, Mann:2011hg, Tarrio:2012xx, langley2020quantum}. 

The variation of the action \eqref{model} around a solution to the field equations reduces to a boundary term,
\begin{equation}
    \delta S_0 = \int d^3x\sqrt{-\gamma}\big((K_{\alpha\beta}-K \gamma_{\alpha\beta})\delta\gamma^{\alpha\beta}-e^{\lambda\phi}n^{\alpha}F_{\alpha\beta}\,\delta A^{\beta} - n^{\alpha}\nabla_{\alpha}\phi\,\delta\phi\big),
    \label{bareactionvar}
\end{equation}
where $n^{\alpha}$ refers to the outward directed unit normal at the boundary. 
As usual, regularisation and renormalisation are needed in order to make sense of this expression. For regularisation, we place the boundary at a large but finite value of the radial coordinate $r$ and introduce boundary counterterms to cancel divergences that would otherwise appear as the boundary regulator is taken to infinity. A number of possible counterterms present themselves, but, as we will see, there is a simple construction that removes all divergences associated with the field variations in \eqref{bareactionvar}. Furthermore, the resulting finite stress tensor complex has precisely the form expected for a general $z$ Lifshitz fluid. The remainder of this subsection draws heavily on reference \cite{Ross:2009ar}, where holographic renormalisation was carried out for gravity coupled to a massive vector field in asymptotically Lifshitz spacetime. 

The model at hand instead has a massless vector field that diverges at the boundary and a logarithmically running dilaton field but the shift symmetry \eqref{shiftsymmetry} suggests a way around that. We can construct a shift-invariant scalar combination of the fields, which takes a constant value, independent of $r$, when evaluated on the pure Lifshitz solution,
\be
e^{\lambda\phi}A_\mu A^\mu\Big\vert_{\eqref{lifshitzbackground}} = -\,\frac{2(z-1)}{(z+2)}.
\ee
In asymptotically Lifshitz spacetime any function of this scalar will contribute at the same order in $r$ to the boundary action and we are led to consider a counterterm action of the form,
\be
S_{c.t.} =\int d^3x\sqrt{-\gamma}\big(-4+f(e^{\lambda\phi}A^2)\big).
\ee
where $f$ is some, as yet, undetermined function. 
The counterterm involving the vector field explicitly breaks the gauge symmetry of the bulk theory. As discussed in Section~\ref{bulktheory}, this is allowed because the only role of $A_\mu$ is to provide the background source that supports the asymptotically Lifshitz geometry and it does not couple directly to matter fields. It can in fact be viewed as an advantage that the gauge symmetry is broken as there is then no question of a conserved $U(1)$ charge in the dual field theory. Such a charge can easily be introduced via an additional bulk gauge field as in \cite{Tarr_o_2011}. 

The first term in the counterterm action (involving $-4\sqrt{-\gamma}$) is a standard counterterm in AdS gravity. With it in place, the variation in \eqref{bareactionvar} involving $\delta\gamma^{tt}$ vanishes for the Lifshitz background \eqref{lifshitzmetric}. This restricts the function in the matter counterterm to be of the form $f(-e^{\lambda\phi}A^2)=\xi \sqrt{-e^{\lambda\phi}A^2}$, with some constant $\xi$. The argument is the same as in \cite{Ross:2009ar} for the massive vector theory, {\it i.e.} the matter counterterm should not contribute at all to the $\delta\gamma^{tt}$ variation and this is ensured if the factor of $\sqrt{-\gamma_{tt}}$ that comes from the $\sqrt{-\gamma}$ prefactor is cancelled by a factor $\sqrt{-\gamma^{tt}}$ from $\sqrt{-e^{\lambda\phi}A_\mu A_\nu \gamma^{\mu\nu}}$. By this simple argument the counterterm action has been determined up to a single constant,
\be\label{ct_action}
S_{c.t.} =\int d^3x\sqrt{-\gamma}\big(-4+\xi\sqrt{-e^{\lambda\phi}A^2}\big),
\ee
and the variation of the full action $S=S_0+S_{c.t.}$ can be written,
\begin{equation}
    \delta S = \int d^3x\Big(s_{\alpha\beta}\,\delta \gamma^{\alpha\beta} + s_{\alpha}\,\delta A^{\alpha} + s\,\delta\phi\Big),
    \label{actionvar}
\end{equation}
with,
\begin{align}
    s_{\alpha\beta} &= \sqrt{-\gamma}\Big(K_{\alpha\beta}-K \gamma_{\alpha\beta} + 2\gamma_{\alpha\beta} - \frac{\xi e^{\lambda\phi}}{2\sqrt{-e^{\lambda\phi}A^2}}(A_{\alpha}A_{\beta} - A^2\gamma_{\alpha\beta}\big)\Big) \label{sab},\\
    s_{\alpha} &= \sqrt{-\gamma}\Big(-e^{\lambda\phi}n^{\beta}F_{\beta\alpha} - \frac{2\xi e^{\lambda\phi}A_{\alpha}}{2\sqrt{-e^{\lambda\phi}A^2}}\Big) \label{sa},\\
    s &= \sqrt{-\gamma}\Big(-n^{\alpha}\nabla_{\alpha}\phi - \frac{\xi\,\lambda e^{\lambda\phi}A^2}{2\sqrt{-e^{\lambda\phi}A^2}}\Big) \label{s}.
\end{align}
Remarkably, all the variations vanish for the Lifshitz background \eqref{lifshitzmetric} for a single value of the free parameter in the boundary counterterm, 
\be
\xi = -\sqrt{2(z-1)(z+2)},
\ee
and no further counterterms are needed for the problem at hand.
For more general backgrounds, including Lifshitz black branes that are asymptotic to the Lifshitz solution, these variations will cancel at leading order and interesting physics resides in subleading terms that do not cancel. For instance, 
requiring the action to be invariant under the Lifshitz scaling \eqref{lifshitzscaling} leads to a Ward identity,
\begin{equation}
    z\, s^t_t + s^x_x + s^y_y + \frac{(z-2)}{2}s^t\,A_t -\sqrt{z-1}\,s = 0,
    \label{lifward}
\end{equation}
which encodes information about the fluid equation of state, as we will see in Section~\ref{sect:thermo} below. 

The Ward identity brings out the different scaling behaviour of temporal and spatial components and this difference should also be manifest in the renormalised stress tensor complex. Indeed, in a non-relativistic system the energy flux $\mathcal{E}^i$ and the momentum density $\mathcal{P}_i$ are independent variables and this appears to be at odds with the symmetry of $s_{\alpha\beta}$ under interchange of $\alpha$ and $\beta$. The resolution involves a subtle issue in holographic renormalisation which arises when a system has background vector fields \cite{Hollands:2005ya,Ross:2009ar}. In this case, we are instructed to introduce orthonormal frame fields $e_\alpha^{(A)}$, with tangent space index $A$, and re-express the metric and vector variations in \eqref{actionvar} in terms of variations involving frame variables,
\begin{equation}\label{framevar}
    \delta S = \int d^3x\Big((-2s^{\alpha}_{\ \beta} + s^{\alpha}A_{\beta})\,e^{\beta}_{(B)} \delta e^{(B)}_{\alpha} + s_B\,\delta A^B + s\,\delta\phi\Big).
\end{equation}
The elements of the boundary stress tensor complex are obtained by considering frame field variations while keeping fixed the dilaton and tangent space components of the vector field \cite{Ross:2009ar},
\begin{align}\label{energies}
\mathcal{E} &= 2s^t_{\ t}-s^t A_t, \qquad\quad \mathcal{E}^i = 2s^i_{\ t} - s^iA_t,\\ 
\mathcal{P}_i &= -2s^t_{\ i} +s^t A_i, \qquad \Pi^i_{\ j} = -2s^i_{\ j} + s^i A_j.
\label{stresscomplex}
\end{align}
In a relativistic theory this prescription ensures a conserved stress tensor. An alternative prescription involving metric variations, keeping fixed spacetime components of a background vector field, would still produce a finite stess tensor but one that does not satisfy the usual conservation law \cite{Hollands:2005ya}. In a non-relativistic theory, however, the tangent space prescription is indispensable. Without the extra terms in \eqref{stresscomplex} that come from the vector field the elements of the stress tensor complex are not guaranteed to be finite, let alone conserved \cite{Ross:2009ar}. We will see this explicitly in the following subsection. When we consider the asymptotic fields of a moving Lifshitz black brane, the metric variation yields a divergence in the energy flux which is exactly cancelled by the contribution from the vector sector.

It is worth noting that for generic $z$ the thermodynamic quantities in the stress tensor complex have different scaling dimensions.  
Straightforward dimensional analysis (see {\it e.g.}\ \cite{Ross_2011}) gives the following scaling behavior in $d_s$ spatial dimensions,
\be
\mathcal{E}' =\Lambda^{-d_s-z}\,\mathcal{E}\,,\quad 
{\mathcal{E}'}^i =\Lambda^{1-d_s-2z}\,\mathcal{E}^i\,,\quad
\mathcal{P}_i' =\Lambda^{-1-d_s}\,\mathcal{P}_i\,,\quad
P' = \Lambda^{-d_s-z}\,P\,.
\label{scalingzz}
\ee

\subsection{The asymptotic solution}
We now wish to consider general backgrounds that approach the Lifshitz solution as $r\rightarrow\infty$. While the metric and gauge field ansatz \eqref{metric} and \eqref{gaugefield} is particularly well suited for deriving conserved Noether charges as in Section~\ref{noetherdef}, it is less convenient for evaluating the stress tensor complex. For the asymptotic analysis we instead adopt a set of orthonormal frame fields,

\begin{align}
\label{e0}
    &e^{(0)} = r^z\,H_1(r)\, dt + r\, v_{1}(r)\, dy,\\ 
   \label{eR}
    &e^{(R)} = \frac{dr}{r\,H_2(r)} , \\\label{e1}
    &e^{(1)} = r\,H_3(r)\, dx,\\ \label{e2}
    &e^{(2)} = r^z\,v_{2}(r)\, dt + r\,H_4(r)\, dy,
\end{align}
and orient the frame field $e^{(0)}$ parallel to the vector field,
$A_\mu= A_0(r)\, e_\mu^{(0)}$, with
\begin{equation}
A_0(r) = \sqrt{\frac{2(z-1)}{(z+2)}}\frac{r^2}{r_h^2}\>a_0(r).   
\end{equation}
Fixing the radial gauge $F_3(r)=1$ amounts to setting $H_3(r)=1$.

In order for the space-time to be asymptotically Lifshitz, we  require that $H_1(r), H_2(r), H_4(r)$, and $a_0(r)$ tend to unity as $r \rightarrow \infty$, while $v_1(r)$ and $r^{1-z}\,v_2(r)$ tend to zero.
With these assumptions, the field equations for a Lifshitz black brane with linear momentum reduce to a set of coupled linear ordinary differential equations for small deviations from the Lifshitz background in the asymptotic region. 

The linearised field equations corresponding to the $ty$ component of Einstein's equations and the $y$ component of Maxwell's equations decouple from the rest of the linearised equations and can be solved separately for $v_1$ and $v_2$. The general solution of the resulting eigenvalue problem consists of four eigenmodes,
\begin{align}
\label{asympv1}
&v_1(r) = c_{1y}\,r^{z-1} + \frac{c_{2y}}{r^3}+\frac{z+2}{3z}\frac{c_{3y}}{r^{2z+1}}\\ \label{asympv2}
&v_2(r) = \frac{c_{4y}}{r^{z-1}}+\frac{c_{2y}}{r^3}+\frac{c_{3y}}{r^{2z+1}},
\end{align}
The $c_{1y}$ and $c_{4y}$ modes are non-renormalisable source modes while $c_{2y}$ and $c_{3y}$ will appear in the momentum density and energy flux, respectively, in the renormalised stress energy complex \cite{Ross_2011}. 

The remaining linearised Einstein, Maxwell and dilaton equations can be solved in a straightforward manner along the lines of \cite{Keranen:2012mx}. The full linearised solution has eigenmodes that go as $r^{-z-2}$, $r^{\frac{1}{2}(-z-2\pm \sqrt{(z+2)(9z+10)})}$ as well as $r^0$. A general holographic analysis of asymptotically Lifshitz space-times  (\cite{Ross:2009ar, Baggio:2011cp, Zingg:2011cw, Mann:2011hg}) shows that a $r^{-z-2}$ mode carries finite energy and is accordingly of primary interest for the analysis that follows. The remaining eigenmodes include a growing mode that would disrupt the asymptotic Lifshitz geometry and must therefore be absent in a physical solution and a mode that falls off at a faster rate than $r^{-z-2}$ and can thus  be ignored in our asymptotic analysis. Finally, we also leave out the constant mode which can be interpreted as a non-normalisable source mode. 

We find that the finite energy mode takes the following form, 
\begin{align}
\label{asympH1}
&H_1 = 1 + \frac{c_1}{2}\, r^{-z-2},\\ \label{asmypH2}
&H_2 = 1 + \frac{c_2}{2}\, r^{-z-2},\\ \label{asympH3}
&H_3 = 1\\ \label{asympH4}
&H_4 = 1 + \frac{c_3}{2}\, r^{-z-2},\\ \label{asympa0}
&a_0 = 1+c_4\, r^{-z-2},\\ \label{asympPhi}
&\phi = 2\sqrt{z-1}\Big(\log\big(\frac{r}{r_h}\big)+ \frac{c_5}{2}\,r^{-z-2}\Big),
\end{align}
with the following relations between the various constants,
\begin{equation}
    c_2 = c_1 + c_3, \qquad c_5 = \frac{c_3}{2}.
    \label{relation1}
\end{equation}
Radial conservation of the charge \eqref{q3}, associated with the shift symmetry of the dilaton field, gives an additional relation,
\begin{equation}
    c_4 = \frac{c_1}{2} -\frac{(z-2)}{4}c_3,
    \label{q3rel}
\end{equation}
leaving us with two independent constants, $c_1$ and $c_3$.

Plugging in the asymptotic solution \eqref{asympH1}-\eqref{asympPhi} into the renormalised stress tensor complex \eqref{stresscomplex} and eliminating constants using \eqref{relation1} and \eqref{q3rel} gives the following  expression
\begin{align}
\label{movE}
&\mathcal{E} = -2c_1 +zc_3 ,\\ \label{movEx}
&\mathcal{E}^x = 0,\\ \label{movEy}
&\mathcal{E}^y = -2(z-1)c_{3y},\\ \label{movPx}
&\Pi^x_{\ x} = zc_1-\frac{1}{2}(z^2-z-2)c_3,\\ \label{movPy}
&\Pi^y_{\ y} = zc_1-\frac{1}{2}(z^2+z+2)c_3,\\ \label{movmomden}
&\mathcal{P}_y = -2(z-1)c_{2y}.
\end{align}
The modes $c_{2y}$ and $c_{3y}$ appear only in the off-diagonal terms of the stress tensor complex and can be interpreted as the momentum density $\rho v$ and energy flux 
$(\mathcal{E}{+}P)v$ respectively. While they cannot be related with the other modes in the absence of the full bulk solution, it appears safe to make this statement, as they show the correct scaling dimensions as required by \eqref{scalingzz}.

In this calculation it was essential to include the vector 
field contribution to the stress tensor complex in \eqref{energies}-\eqref{stresscomplex}. The energy flux, given by
$\mathcal{E}^i = 2s^i_{\ t} - s^iA_t$,
is divergent if only the contribution from the $s^i_{\ t}$ term is included but when both terms are present the divergence is cancelled, leaving behind a finite result for $\mathcal{E}^i$. Moreover, the resulting finite energy flux \eqref{movEy} and the momentum density \eqref{movPy} are independent in that they involve different eigenmodes of the linearised solution.

\section{Thermodynamics}\label{sect:thermo}

Having constructed a finite renormalised stress energy complex, we can proceed to study the thermodynamics of the fluid described by the above holographic dual. We expect to find a perfect Lifshitz fluid, whose description is invariant
under time and space translations as well as rotations, but without boost symmetry. 

The thermodynamic relations that we wish to establish require us to supplement the linearised solution in the asymptotic region with near-horizon data. This can be done via the Noether charges that we obtained earlier, even in the absence of an exact (or numerical) solution that interpolates between the near-horizon and asymptotic regions.

\subsection{The near-horizon solution}
Using similar arguments to those made in \cite{Hartong_2016}, we find that an appropriate near-horizon Taylor series expansion of the various metric and gauge fields in \eqref{metric} and \eqref{gaugefield} is given by,
\begin{align}
F_1 &= f_1 \frac{(r-r_h)}{r_h} + \ldots\\
F_2 &= h_1 \frac{(r-r_h)}{r_h}+ \ldots\\
F_3 &= 1\\
F_4 &= p_0 + p_1 \frac{(r-r_h)}{r_h}+ \ldots\\
N &= n_0 + n_1 \frac{(r-r_h)}{r_h}+\ldots\\
G_1 &= g_1 \frac{(r-r_h)}{r_h}+\ldots\\
G_2 &= m_0 + m_1 \frac{(r-r_h)}{r_h}+\ldots\\
\phi &= l_0 + l_1\frac{(r-r_h)}{r_h}+\ldots,
\end{align}
\label{nhexp}
where, as before, we are working in the $F_3 = 1$ gauge.
The location of the horizon $r_h$ is where $g^{rr}$ vanishes. This means that $F_2$ will have a first order zero at $r_h$. Also, regularity of the metric in Eddington-Finklestein coordinates requires $F_1$ to have a first order zero at $r_h$ as well. Finally, $G_1$ should also have a first order zero at $r_h$, but the remaining fields are regular there. 

The near horizon metric is then given by,
\begin{equation}
ds^2 = -f_1\rho r_{h}^{2z-1}dt^2+\frac{d\rho^2}{r_h h_1 \rho} + r_{h}^2dx^2 + p_0r_{h}^2\big(dy+N(r_h)r_{h}^{z-1}dt\big)^2,
\end{equation}
where $\rho = r-r_h$.
Making a change of coordinates,
\begin{equation}
\tilde{\rho}^2 = \frac{4\rho}{r_h h_1}, \: \tilde{t} = (f_1h_1)^{\frac{1}{2}}\frac{r_h^zt}{2},
\end{equation}
the metric takes the following form,
\begin{equation}
ds^2 = -\tilde{\rho}^2d\tilde{t}^2 + d\tilde{\rho}^2 + r_h^2dx^2 + p_0r_{h}^2\big(dy+N(r_h)r_{h}^{z-1}dt\big)^2.
\label{nhmet}
\end{equation}
After going to Euclidean coordinates, the Hawking temperature is found to be $T = \frac{r_h^z}{4\pi}(f_1h_1)^{\frac{1}{2}}$. Meanwhile (using units where $16\pi G_N = 1$),
the entropy is given by $s = \frac{A}{4G_N} = 4\pi p_0^{\frac{1}{2}}r_h^2$, and we have
\begin{equation}
T s = (p_0f_1h_1)^{\frac{1}{2}}r_h^{z+2}.
\label{ts}
\end{equation}
In the following subsections we consider in turn the static and the moving black branes and compare this expression to the Noether charge(s) obtained in the near-horizon region. Via the conservation of the Noether charges, this can then be related to the asymptotic solution and the stress tensor complex evaluated at the boundary.

The radial gauge conditions $F_3(r)=1$ and $H_3(r)=1$ ensure that we can identify the radial variable $r$ in our near-horizon solution with that in our asymptotic solutions and then there is a straightforward mapping between the frame fields defined in \eqref{e0}-\eqref{e2} and coordinate fields \eqref{metric}, which we can use to evaluate the Noether charges at the boundary. 

\subsection{Static Lifshitz black brane}
As a warm-up exercise, let us first study thermodynamics of a static Lifshitz brane with general $z$ given by the static limit of \eqref{metric} by evaluating the renormalised stress tensor at the boundary and using the method of Noether charges.

For a static black brane we have $c_3=0$ and the stress tensor complex in \eqref{movE}-\eqref{movmomden} reduces to
\begin{align}
\label{statE}
&\mathcal{E} = -2c_1,\\ \label{statEx}
&\mathcal{E}^x = 0,\\ \label{statEy}
&\mathcal{E}^y = 0,\\ \label{statPx}
&\Pi^x_{x} = zc_1,\\ \label{statPy}
&\Pi^y_{y} = zc_1,\\ \label{statmomden}
&\mathcal{P}_{y} = 0.
\end{align}
One can easily verify that the scaling Lifshitz Ward identity \eqref{lifward} is satisfied. In fact, $s_{\alpha}$ and $s$ vanish trivially for the static brane. Identifying $-2c_1$ and $-z\,c_1$ as the energy and pressure $P$, respectively, we obtain
\begin{equation}\label{stateeq}
    z\mathcal{E} = 2P,
\end{equation}
which is the well-known equation of state for a static perfect Lifshitz fluid.

Plugging in the near horizon metric anstaz defined in \eqref{nhexp} into our Noether charges obtained in \eqref{noetherdef}, we find that $Q_{\beta}$ vanishes and  $Q_{\alpha}$ reduces to 
\begin{equation}
Q_{\alpha} =  T s,
\label{statnoethernh}
\end{equation}
with $T$ the Hawking temperature and $s$ the entropy density of the static brane. 
Plugging in the asymptotic metric ansatz \eqref{asympH1}-\eqref{asympPhi} into the non-vanishing Noether charge gives
\begin{equation}
    Q_{\alpha} = -(z+2)c_1.
    \label{statnoetherbndry}
\end{equation}
Equating the right hand sides of \eqref{statnoethernh} and \eqref{statnoetherbndry} and identifying $\mathcal{E}=-2c_1$ from \eqref{statE}, we get precisely the Gibbs-Duhem relation expected for a Lifshitz fluid,
\begin{equation}
    \mathcal{E} = \frac{2}{z+2}Ts,
\end{equation}
or, equivalently,
\be
\mathcal{E}+P=Ts .
\ee

\subsection{Moving Lifshitz black brane}

Let us now proceed to repeat the above exercise for the case of interest, that is, a Lifshitz black brane with linear momentum along one of the transverse directions (which we chose to be the $y$-direction without loss of generality).

For a perfect fluid moving in the y-direction
the pressure is identified via $\Pi^x_{\ x}=-P$ 
while $\Pi^y_{\ y} = -P - \rho\, v^2$ is the combined stress due to pressure and momentum flow along the $y$ direction. The stress tensor complex for the moving Lifshitz black brane was obtained in \eqref{movE}-\eqref{movmomden} and we can read off the energy density, pressure, and momentum flux,
\be\label{perfectfluid}
\mathcal{E}=-2c_1+z\,c_3, \quad P=-z\, c_1+\frac{1}{2}(z^2-z-2)c_3,\quad \rho\, v^2=(z+2)c_3. 
\ee
Using the linearised solution we can easily verify that the Lifshitz Ward identity \eqref{lifward} is indeed satisfied and gives us the equation of state for a perfect Lifshitz fluid in motion,
\begin{equation}
    z\mathcal{E} -\rho v^2 = 2P,
    \label{eoslifz}
\end{equation}
as hypothesised in \cite{de_Boer_2018,de_Boer_2020}. As discussed above, the modes $c_{2y}$ and $c_{3y}$ appear only in the off-diagonal terms of the stress tensor complex. 
Although they cannot be related to the other modes in the absence of an interpolating solution, they do have the right scaling to be interpreted as the momentum density $\rho v$ and energy flux $(\mathcal{E}{+}P)v$, respectively.

Plugging in the near horizon metric anstaz defined in \eqref{nhexp} into our Noether charges $Q_\alpha$ and $Q_\beta$ obtained in \eqref{noetherdef}, we find that a simple linear combination of the two in fact gives us \eqref{ts}. Specifically,
\begin{equation}
Q_{\alpha} - \frac{3}{2}Q_{\beta} = T s.
\label{noethernh}
\end{equation}

Plugging in the asymptotic metric ansatz \eqref{asympH1}-\eqref{asympPhi} into the same Noether charges gives
\begin{align}
Q_{\alpha} &= \frac{(z+2)}{2}\,(-2c_1 + z\,c_3),\\
Q_{\beta} &= (z+2)c_3=\rho\, v^2.
\end{align}
Now use \eqref{perfectfluid} and the equation of state \eqref{eoslifz} to express the Noether charges in terms of fluid variables, 
\begin{align}
Q_{\alpha} &= \mathcal{E}+P+ \frac{1}{2}\rho\, v^2,\\
Q_{\beta} &=  \rho\, v^2,
\end{align}
and it immediately follows that the combination of Noether charges in \eqref{noethernh} gives the Gibbs-Duhem relation for a moving Lifshitz perfect fluid,
\begin{equation}
   \mathcal{E} + P = Ts + \rho v^2.
\label{noetherbndy}
\end{equation}
Interpreting the various boundary data as field theory sources in this manner, we find that the velocity of the fluid $v$ appears as the chemical potential conjugate to the momentum density $\rho v$ in accordance with \cite{Hartong_2016,de_Boer_2018}.

Finally, it is interesting to consider the special value of $z=2$, for which the Lifshitz scaling symmetry is consistent with Galilean boost symmetry. In this case, the energy density, pressure, and momentum flux in \eqref{perfectfluid} reduce to 
\be\label{z2fluid}
\mathcal{E}=-2c_1+2c_3, \quad P=-2 c_1,\quad \rho\, v^2=4c_3, 
\ee
The energy density can be written as the sum of an internal energy density and a kinetic energy density, $\mathcal{E}=\mathcal{E}_0+\frac12 \rho\, v^2$. In a Galilean boost invariant system $\mathcal{E}_0$ is simply the energy density in the rest frame of the fluid and the pressure, $P=\mathcal{E}_0$, is frame independent. For general $z\neq 2$, these last two statements do not hold.

\section{Conclusion and Future outlook}

In this paper, we have investigated the holographic dual of a perfect Lifshitz fluid with an arbitrary dynamical exponent, $z$, moving with a velocity, $v$. The moving black brane  we have considered is a solution of the Einstein-Maxwell-Dilaton theory, and is obtained by constructing a new class of metrics that correspond to a black brane having a linear momentum, rather than boosting a static black brane. 

By evaluating the Noether charges associated with various scaling symmetries of our metric, and equating their values at the horizon and boundary (as they are radially conserved quantities), we were able to obtain the expected equation of state for the perfect Lifshitz fluid with an arbitrary dynamical exponent, $z$ (\cite{de_Boer_2018}). Moreover we note that the velocity indeed appears as a chemical potential, conjugate to the dual momentum density (\cite{de_Boer_2018}, \cite{Hartong_2016}).

By solving the linearised perturbations of the Einstein, Maxwell and dilaton equations, we found the asymptotic solutions of the metric and gauge fields, and used these to construct the boundary stress tensor of the fluid. By working in an orthonormal frame and using counterterms similar to those proposed in \cite{Ross:2009ar}, we were able to construct the renormalised stress tensor complex and read off the various thermodynamic variables. The result has the expected form for a non-relativistic fluid, and satisfies the Ward identity associated with Lifshitz scaling. 

There are several interesting questions to be investigated. A straightforward extension of the present work would be to study the gravitational dual of charged perfect Lifshitz fluids via the addition of other $U(1)$ fields to the theory, for which the boundary theory that one might need to study would be that of non-relativistic electrodynamics coupled to torsionless Newton-Cartan theory, as explored in \cite{festuccia2016symmetries}. Another extension would be to carry out the analysis performed in this paper on theories with a non-trivial coupling between the dilaton and cosmological constant, that is, moving hyperscale-violating Lifshitz geometries, and study the resulting thermodynamics.

Lifshitz holography is still a relatively unexplored area as compared to it's relativistic counterpart. Following the work done in \cite{de_Boer_2020} where the complete first-order energy-momentum tensor in curved space-time for a fluid without boost symmetry was computed, one can now consider trying to find the hydrodynamic modes of Lifshitz fluids from quasinormal modes and compare the results.  In this study, we have concentrated on perfect fluids without impurities or lattice effects which break translational invariance. Physical systems that break boost invariance, however, often also exhibit other spontaneous symmetry breaking patterns. While the main objective of the work carried out in \cite{Armas:2020mpr} was to provide a general hydrodynamic framework for fluids that break boost symmetry, it also offers a controlled framework for studying other patterns of symmetry breaking. It would be of interest to study Lifshitz gravity duals that could model this behaviour, in order to better understand their microscopics. With the current holographic model for a moving Lifshitz fluid established, one can also proceed to calculate two point correlation functions related to transport and other quantities of interest from a gauge-gravity perspective. 

\noindent\underline{Acknowledgements:} We would like to thank Valentina Giangreco M. Puletti, Jelle Hartong, Niels Obers, Nick Poovuttikul, and Watse Sybesma for useful discussions and input. This research was supported in part by the Icelandic Research Fund under grant 195970-052 and by the University of Iceland Research Fund.

\bibliographystyle{utphys}

\bibliography{boostedbhbiblio}

\end{document}